\documentclass[a4paper,
               keeplastbox,   
               ]{jacow}
\usepackage{pdfpages,multirow,ragged2e} %
\makeatletter%
	\ifboolexpr{bool{xetex}}
	 {\renewcommand{\Gin@extensions}{.pdf,%
	                    .png,.jpg,.bmp,.pict,.tif,.psd,.mac,.sga,.tga,.gif,%
	                    .eps,.ps,%
	                    }}{}
\makeatother

\ifboolexpr{bool{xetex} or bool{luatex}} 
 {}                                      
 {\usepackage[utf8]{inputenc}}           

\usepackage[english]{babel}

\ifboolexpr{bool{jacowbiblatex}}%
 {%
  \addbibresource{jacow-test.bib}
  \addbibresource{biblatex-examples.bib}
 }{}
\begin{document}

\title{Loss maps along the \NoCaseChange{ThomX} transfer line\\ and the ring first turn}

\author{A. Moutardier\thanks{moutardier@ijclab.in2p3.fr}, C. Bruni, I. Chaikovska, S. Chancé, N. Delerue, E. E. Ergenlik, V. Kubytskyi,\\ H. Monard, Université Paris-Saclay, CNRS/IN2P3, IJCLab, Orsay, France}
	
\maketitle

\begin{abstract}

We report on studies of the loss maps for particles travelling from the end of the ThomX's linac along the transfer line to the end of the ring first turn in preparation of the machine commissioning. ThomX is a 50-MeV-electron accelerator prototype which will use Compton backscattering to generate a high flux of hard X-rays.
The accelerator tracking code MadX is used to simulate electrons' propagation and compute losses.
These maps may be projected at any localisation along the bunch path or plotted along the bunch path. This information is particularly relevant at the locations of the monitoring devices (screens, position monitors,...) where loss predictions will be compared with measurements.

\end{abstract}

\section{\NoCaseChange{ThomX}}

ThomX is a 50-MeV-electron accelerator using Compton backscattering to generate a high X-ray flux.

The beam is produced at the photocathode and accelerated in a \SI{3}{GHz} accelerating section before going into the transfer line (TL). 
This line may be used in two configurations, one straight after the linac with diagnostics (OTR screens, faraday cup,..., see \cite{diag_op,diag_e-}
) to characterise the beam, and an other one bent at \SI{90}{\degree} by two magnets to shape the beam for the ring injection. This second configuration includes a dispersive line to characterise the beam energy spread.
In the TL there are three diagnostic stations (SST) used to visualise the beam shape with several screens and a camera each.

\begin{figure}[!h]
	\centering
	\includegraphics*[width=0.9\columnwidth, trim=19 0 0 0]{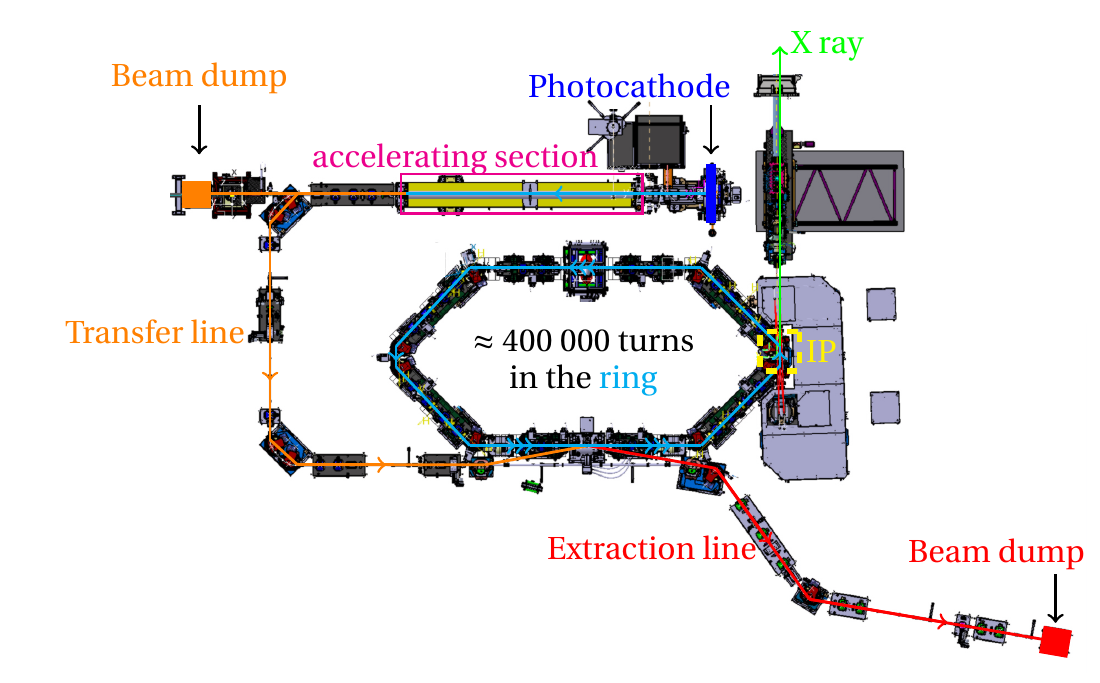}
	\caption{ThomX overview. The electron beam is generated in the photo-cathode (in {blue}) then accelerated within an accelerating section (in {pink}). The transfer line (in {orange}) is divided in two configurations, one that goes straight to a beam dump for initial settings and an other one that is bent towards a dispersive line and then to the ring (in {cyan}). Each time the beam passes through the Interaction Point (IP in {yellow}) it interacts with a laser beam to generate X-rays (in {green}) by Compton backscattering. After about 400~000~turns (\SI{20}{\milli\second}) in the ring, the beam will be extracted through the extraction line (in {red}) to a second beam dump.}
	\label{fig:ThomX}
\end{figure}

Once the beam is ready to be injected in the ring, an injection dipole propagates it through a septum to the ring.
A fast kicker in the ring is used to correct the injection angle and allow the beam to propagate on the optimal ring orbit. 
Each time the beam passes through the Interaction Point \hbox{(see Fig.~\ref{fig:ThomX})} it encounters a laser beam stored in a Fabry-Perot cavity. Some electrons may transfer part of their momentum to \SI{1}{eV} photons coming backward and increase the photon energy up to \SI{70}{keV} (hard X-rays).
To feed the Fabry-Perot cavity a fibre laser need to be close to the beam pipe but it is sensitive to radiations induced by electron beam losses.

The machine commissioning is expected to start soon. In preparation we study particle losses.

\section{MadX Tracking Code}

Beam losses are predicted by tracking particles with the accelerator tracking code MadX~\cite{MadX} - developed by CERN - and checking when they are outside the beam pipe aperture.

Moreover the tracking module allows one to recover where particle losses occurred. These lost particles are no longer tracked.

\section{Limiting Aperture and Beam Parameters}

\begin{figure}[!htb]
   \centering
   \includegraphics*[width=0.75\columnwidth]{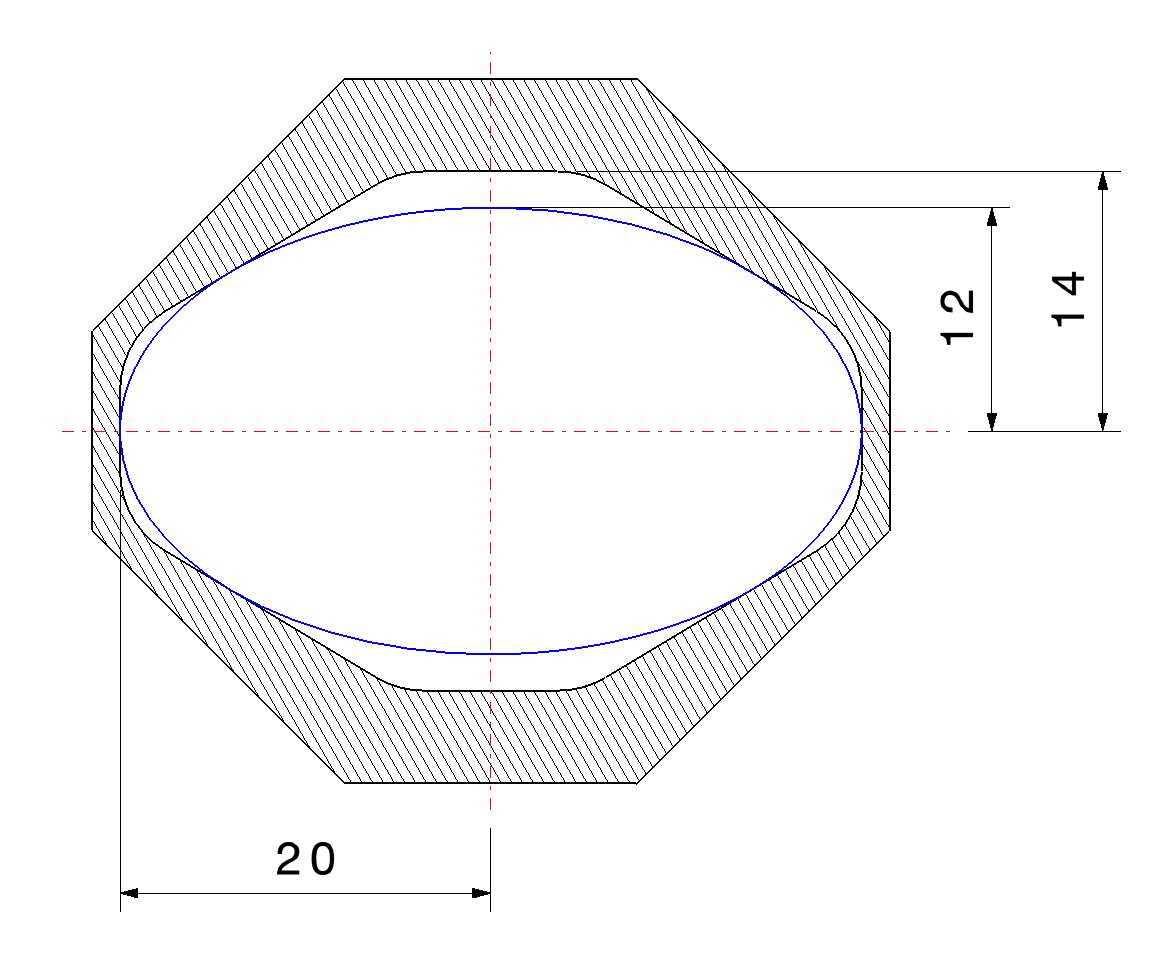}
   \caption{Drawing of the ring's beam pipe from the front. Real pipe is shown with hashes. Aperture shape used in MadX ring's representation is in blue line. Units : \SI{}{\milli\meter}.}
   \label{fig:ring_pipe_shape}
\end{figure}

ThomX's TL and ring has been implemented in MadX and the aperture of each element has been defined based on mechanical data. In the TL the pipe is circular with a radius of \SI{17.5}{\milli\meter} whereas in the ring its shape is more complicated (see Fig.~\ref{fig:ring_pipe_shape}) .

The ring's shape was simplified by taking the maximum ellipse completely fitting in the ring pipe. This ellipse has a semi-minor axe of \SI{12}{\milli\meter} and semi-major axe of \SI{20}{\milli\meter} as shown on Fig.~\ref{fig:ring_pipe_shape} in blue line.
The beam losses are slightly overestimated in the ring.

\section{Simulation Parameters}
\subsection{Lattice}

Some simulations have been done to predict the beam parameters at the exit of the accelerating section. 

The emittance is $\epsilon$ = \SI{5.0E-8}{\meter\radian} in both plan and Twiss parameters are:
\begin{itemize}
    \item $\beta_x$ = $\beta_y$ = \SI{43.2}{\meter}
    \item $\alpha_x$ = $\alpha_y$ = \SI{11.0}{}
\end{itemize}

Hence, the nominal beam size is $\sigma = \sqrt{\epsilon\beta} = \SI{1.5}{\milli\meter}$.

MadX was also used to calculate a periodic beam solution in the ring and the TL quadrupoles' strength to match beam parameters at the entrance of the TL with this periodic solution. 
The lattice used there is slightly different from the one presented in the project's TDR~\cite{TDR}.




\subsection{Random Selection of Particles}


To simulate relevant particles it was chosen to select them within the phase space of the beam described above but with an emittance $\epsilon$ multiplied by $k^2$ such that the beam size is $k\times\sigma$.

For each particle, the code selects randomly (x,$p_x$) - and then (y,$p_y$) - with a uniform random distribution until it is within the beam ellipse. 
An additional \SI{35}{mm} long square cut on x-y plan is added as the particles outside this square cut are well outside the beam pipe.
The selected particles are then propagated, using MadX, from the beginning of the transfer line to the end of the first ring turn. 

\section{Loss Map}
\subsection{Losses Along ThomX}

\begin{figure*}[!bth]
    \centering
    \includegraphics*[width=0.95\textwidth]{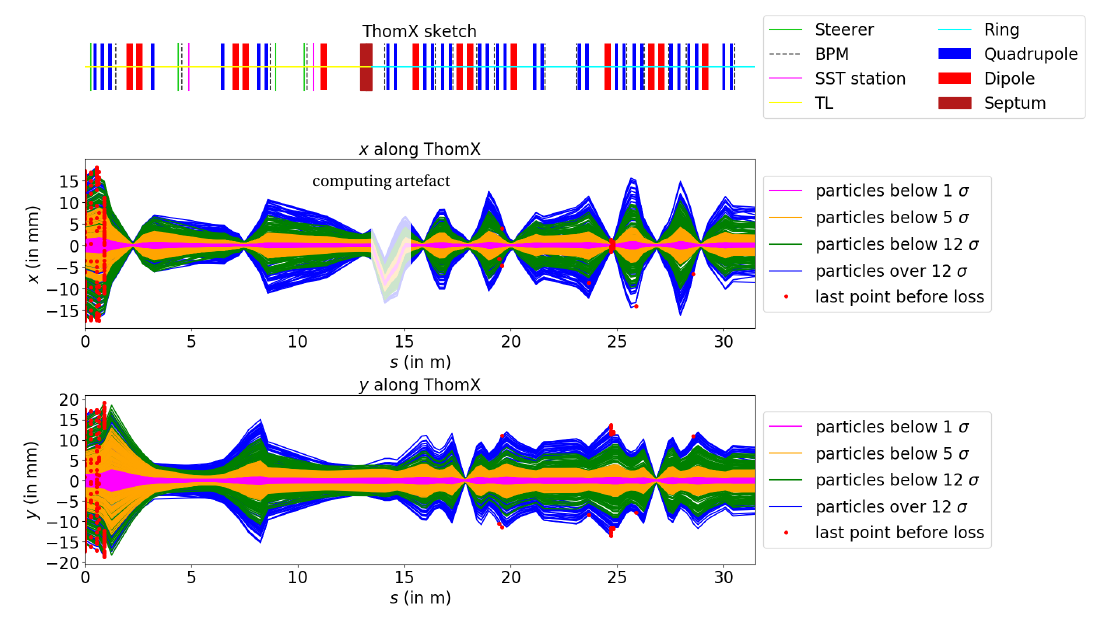}
    \caption{Losses along ThomX. 
    The upper part represent the ThomX line from the TL to the end of the first ring turn.
    The lower part is plots of tracking of particles in x (at the middle) and y (at the bottom) plan along the accelerator.}
    \label{fig:loss_along_ThomX}
\end{figure*}

The localisation of losses may be represented along the accelerator (see Fig.~\ref{fig:loss_along_ThomX}). The computing artefact between s = \SI{13.5}{\meter} and s = \SI{15}{\meter} comes from a change of frame at the entrance of the ring to take into account the off-axes propagation in some elements between the septum and the kicker during the injection. Hence, in the white box, particles are not represented in the beam reference frame, but in the ring reference frame.


For the Fig.~\ref{fig:loss_along_ThomX}, it was chosen to select 4$\times$100 particles within 4 beams characterised by 1, 5, 12 and 20~$\sigma$  and to represent their trajectory in x and y plan along the accelerator. The last localisation before a particle's loss is represented with a red dot. 

For both 1 and 6~$\sigma$ cases there are not any losses of particles. If the beam is as predicted by simulations, almost no particles' loss will occur on the TL and the ring first turn as only a few electrons propagate outside 5~$\sigma$.

The case 12~$\sigma$ corresponds to the case where particles at the border of the beam are exactly at the border of the pipe. Depending on their localisation and transverse momentum, some particles may be kept while others are quickly lost but no losses are encountered after the first dipole of the TL. Hence, even a very huge beam will create only a few losses. Moreover, those losses are located at the exit of the accelerating section far away from the Interaction Point and its fragile optical fibre.

All the losses encountered after the first dipole of the TL come from particles within the 20~$\sigma$ beam but outside the 12~$\sigma$ one. As the 12~$\sigma$ beam represents the maximum beam size in x and y plan, increasing $k$ over 12 is equivalent to increasing the transverse momentum.
Hence the losses encountered in the ring come from particles with high px or py values at the beginning of the TL. 
A physical measurement of beam parameters at the beginning of the TL before the first beam injection will permit to check px and py values and evaluate the risk of losses in the ring. 
Moreover, it seems that those potential losses are mostly located after the Interaction Point - which is at \SI{18}{\meter} - hence they are not directed toward the Fabry-Perot cavity.

\subsection{Projected Map of Losses}

An other interesting way to represent losses is to represent each particle at one localisation - for example at the beginning of the TL or at the positions of the SST stations - and to project the position of the particle losses on this plot. 

\begin{figure*}[!bth]
    \centering
    \includegraphics*[width=0.75\textwidth]{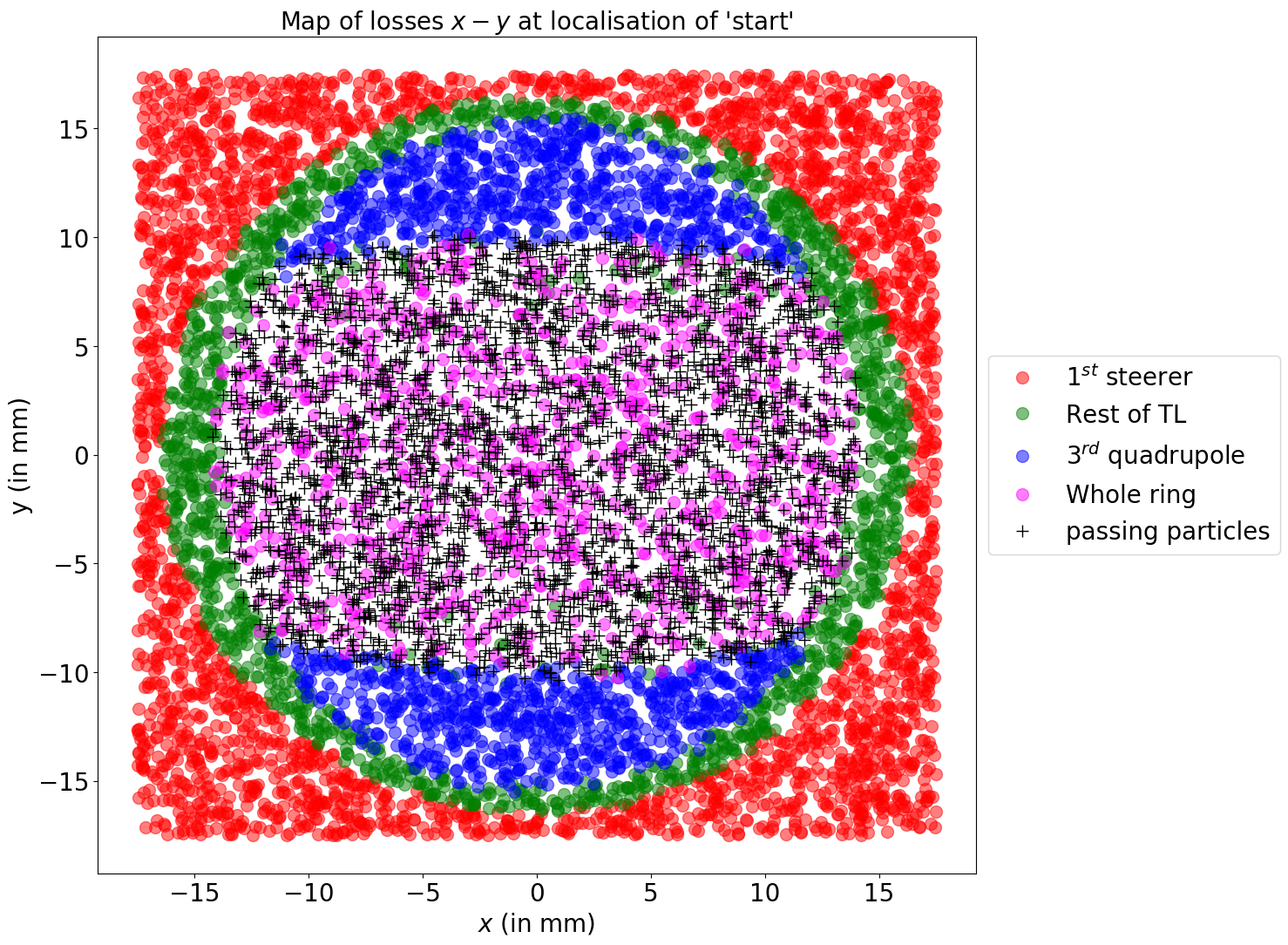}
    \caption{Map of losses along the TL and the first turn in the ring, projected at the end of the accelerating section.}
    \label{fig:loss_map}
\end{figure*}

An example of this projected map of losses on the x-y plan at the beginning of the TL is shown Fig.~\ref{fig:loss_map}.

This time the \SI{e4}{} particles are randomly selected within a 20~$\sigma$ beam only.
One may clearly see the square cut in the x-y plan during the selection of particles and the circular cut at the first element done by MadX.

The shape of losses on the 3rd quadrupole of the TL shows that the beam losses occur at this point because of the large vertical beam size. The same conclusion applies to the last series of red dots before \SI{2}{\meter} on the Fig.~\ref{fig:loss_along_ThomX}.

At the middle of the plot, a window of acceptance parameters can be seen. In this region particles are either particles that will not be lost (black cross) or particles coming from $k~\sigma$ beam, with $k$ greater than 12. 

Finally, knowing the size of the acceptance window at the locations of the SST stations may permit to check the possibility for a real beam to be transmitted until the end of the first ring turn even before trying it.

\section{Conclusion}

Maps of losses along an accelerator and projected maps of losses are efficient ways to predict localisation of losses and allow one to check beam losses even before sending the beam in some critical part of the accelerator.

On ThomX the window of acceptance parameters will be calculated at the localisation of each SST station and a graphical representation of it will be added on beam images on the TL to check the risk of losses before injecting a beam in the ring. Losses investigations must be done with real beam parameters but primary losses simulations show that beam losses are well controlled.

\section{ACKNOWLEDGEMENTS}
THOMX is financed by the French National Research Agency under the EQuipex program ANR-EQPX-51.

%
\ifboolexpr{bool{jacowbiblatex}}%
	{\printbibliography}%

\begin{thebibliography}{9} 
	
%

	\bibitem{diag_op}
	    N. Delerue \emph{et al.},
        \textquotedblleft{Overview of the Optical Diagnostics of the ThomX Compact Compton Source}\textquotedblright,
        presented at the 12th Int. Particle Accelerator Conf. (IPAC’21), Campinas, Brazil, May 2021, paper MOPAB305, this conference.

	\bibitem{diag_e-}
    	N. Delerue \emph{et al.},
        \textquotedblleft{Overview of the Electronic Diagnostics of the Thomx Compact Compton Source}\textquotedblright,
        presented at the 12th Int. Particle Accelerator Conf. (IPAC’21), Campinas, Brazil, May 2021, paper MOPAB306, this conference. 
    
	\bibitem{MadX}
		F. C. Iselin, \emph{The MAD Program (Methodical Accelerator Design): Physical Methods Manual}, Version 8.13, Geneva, Switzerland, Sep. 1994, pp. 1-74; \url{http://mad8.web.cern.ch/mad8/doc/phys_guide.pdf}
	\bibitem{TDR}
		A. Variola, J. Haissinski, A. Loulergue, F. Zomer, \textquotedblleft{ThomX Technical Design Report}\textquotedblright, SOLEIL, Gif-sur-Yvette, France, Rep. in2p3-00971281, Apr. 2014.  


	\end{thebibliography}
	{%
	
	
} 
%
%


\end{document}